\documentstyle[12pt]{article}

\newcommand{\com}[2]{\left[ #1,#2 \right]}               
\newcommand{\acom}[2]{ \{ #1,#2 \} }                     
\newcommand{\resetcounter}{\setcounter{equation}{0}}     
\newcommand{\lagrange}{{\cal L}}

\newcommand{\GG} {{\Delta}}               
\newcommand{\T}  {\theta}
\newcommand{\Tb} {\bar\theta}
\newcommand{\A}  {\alpha}
\newcommand{\B}  {\beta}
\newcommand{\E}  {\varepsilon}

\newcommand{\Sm} {\sigma}

\begin{document}

\thispagestyle{empty}
\begin{titlepage}
\begin{flushright}
HUB--EP--96/35 \\
hep-th/9607216 \\
\end{flushright}
\vspace{0.3cm}
\begin{center}
\Large \bf The Hypermultiplet in N = 2 Superspace
\end{center}
\vspace{0.5cm}
\begin{center}
Ingo \ Gaida$^{\hbox{\footnotesize{1}}}$\\
\vspace{0.3cm}
{\em  Humboldt--Universit\"at zu Berlin, Institut f\"ur Physik,\\
 Invalidenstrasse 110, D--10115 Berlin, Germany}
\end{center}
\vspace{1cm}

\begin{abstract}
\noindent
Global $N=2$ supersymmetry in four dimensions with a 
Fayet-Sohnius hypermultiplet and a complex central charge 
is studied in $N = 2$ superspace. 
It is shown how to construct the complete expansion of
the hypermultiplet with respect to the central charge. In addition
the low-energy effective action is discussed and it is shown
that the `kernel' of the Lagrangian only needs
an integration over a `small' superspace to construct a
supersymmetric action. 
\end{abstract}

\vspace{0.3cm}
\footnotetext[1]{e-mail: gaida@qft2.physik.hu-berlin.de,
                 supported by Cusanuswerk}
\vfill
\end{titlepage}


\setcounter{page}{1}

%
%

\resetcounter

\setcounter{section}{1}

In this letter, a covariant formulation of $N=2$ superspace with 
global $N=2$ supersymmetry and a complex central charge is studied. 
Because of the interesting physical properties of $N=2$ supersymmetric
theories like, for instance, 
electric-magnetic duality [\ref{Montonen_Olive},\ref{Seiberg_Witten}]
they have been extensively studied during the past years
[\ref{Theisen_Klemm},\ref{Louis},\ref{Kachru_Vafa},\ref{Luest}].
The supersymmetry generators $\GG_{A}$ of any supersymmetric theory 
obey a graded Lie algebra
 
\begin{eqnarray}
 \com{\GG_{A}}{\GG_{B}} 
 &=& \GG_{A} \GG_{B} - \ (-)^{ab} \GG_{B} \GG_{A}
\ = \ T_{AB}^{ \ \ \ C} \ \GG_{C}.
\end{eqnarray}

The torsions $T_{AB}^{ \ \ \ C}$ obey certain constraints depending
on the number of supersymmetry generators and the structure
group considered. 
It is possible to perform a change of basis in such a way that the 
supersymmetry generators become derivatives in superspace and the 
momentum operator
becomes the usual partial derivative of four-dimensional Minkowski space.
In this new basis the supersymmetry algebra is

\begin{eqnarray}
\label{graded_lie_algebra2}
 \com{D_{A}}{D_{B}} 
 \ + \ i \ T_{AB}^{ \ \ \ C} \ D_{C} &=& 0
\end{eqnarray}

In $N=2$ supersymmetric theories central charges
can occur in the supersymmetry algebra [\ref{Haag}]. 
The corresponding $N=2$ supersymmetry multiplets may be charged with respect
to this central charge. For instance, the hypermuliplet
[\ref{Fayet},\ref{Sohnius}] carrys central charge, whereas the
vector multiplet does not [\ref{Grimm_Wess_Sohnius}, \ref{Gaida}]. 
In the context of the electric-magnetic duality 
the central charge plays a crucial  r{\^ o}le because the electric and the 
magnetic charges can appear as a complex central charge in the 
$N=2$ supersymmetry algebra [\ref{Witten_Olive}]. 
\\
Hypermultiplets are of special interest in the analysis of 
electric-magnetic duality and non-perturbative dynamics,
because in low-energy effective actions for
$N=2$ supersymmetric Yang-Mills theories the monopoles and dyons are
massive hypermultiplets
[\ref{Seiberg_Witten},\ref{Theisen_Klemm}]. In a 
free field theory they saturate the BPS bound on-shell

\begin{eqnarray}
\label{BPS_bound}
 M^{2} &\geq& | \ Z \ |^{2}.
\end{eqnarray}

Here $M$ denotes the mass of the soliton and $Z$ its central charge. 
These results of $N=2$ supersymmetric gauge theories have been
already applied to string theory [\ref{Louis},\ref{Kachru_Vafa},\ref{Luest}].
Moreover, because black holes are the solitons of quantum gravity,
hypermultiplets also occur in the discussion
of extreme black holes [\ref{Strominger},\ref{Kallosh}].
It is essential in all these considerations that the hypermultiplet is
a $N = 2$ supersymmetric multiplet with spin $\leq 1/2$ describing
electrically and magnetically charged matter in four dimensions.
However, the difficult off-shell structure of the hypermultiplet doesn't 
play any r{\^ o}le. 
\\
The discussion of supersymmetric quantum field theories in the framework
of a superspace formulation is an old and successful approach. 
Although the $N=1$ superspace formulation is established and the general
couplings are well known, things
are  different in theories with $N \geq 2$. For instance, Berkovits 
and Siegel recently discussed superspace effective actions of heterotic
and Type II superstrings [\ref{Siegel}] and found contradictions
with the standard folklore [\ref{Strominger}] that Type II
string loops are counted by just a hypermultiplet.
\\
All these results and considerations motivate trying to
gain a better understanding of $N=2$ superspace, 
the hypermultiplet and the central charge on general grounds.
One goal in the present paper is to give the off-shell
expansion of the hypermultiplet in the presence of a complex central 
charge in a covariant formulation of $N = 2$ superspace.
In 1978 Sohnius wrote down a superfield representation of the
hypermultiplet in the presence of one central charge 
and a remarkable action formula in superspace [\ref{Sohnius}].
In the present paper his approach is the starting point.
It is shown that in the presence of a complex central charge an
additional constraint must hold to keep the minimal field content of
the Fayet-Sohnius hypermultiplet.
In particular it is shown how to construct the complete expansion of
the hypermultiplet with respect to the central charge.
In addition the low-energy effective action is discussed and it is shown
that the `kernel' of the Lagrangian only needs an integration over a `small'
superspace to construct a supersymmetric action. 
\\
The letter is organized as follows:
First the $N=2$ supersymmetry algebra
and $N=2$ superspace is introduced. Then the hypermultiplet
is discussed at the superfield and the component level in great detail.
A discussion of the hypermultiplet\footnote{ The so-called 
`relaxed hypermultiplet' [\ref{Townsend1}] is not considered here.} 
exists also in the harmonic
superspace formalism [\ref{harmonic_superspace}], but this approach is not
considered here. Moreover there exists another approach to consider 
central charge transformations of $N=2$ scalar multiplets [\ref{Lauwers}].
\\
\\
Using the conventions of [\ref{Wess_Bagger}] the
$N=2$ supersymmetry generators 
$\Delta_{A} = \{ P_{m}, Q_{\A}^{i},\bar Q_{\dot\A j}, Z, \bar Z, \}$
obey the following algebra

\begin{eqnarray}
\label{n=2susy_algebra}
 \acom{ Q_{\A}^{i} }{ \bar Q_{\dot\A j} } &=& 2 \ \delta^{i}_{ \ \ j} \
                                 \Sm_{ \ \ \A \dot\A}^{m} \ P_{m} 
\nonumber\\
 \acom{ Q_{\A}^{i} }{ Q_{\B}^{j} }        &=& 2 \ g^{ij} \ \E_{\A\B} \ Z 
\nonumber\\
\acom{ \bar Q_{\dot\A i} }{ \bar Q_{\dot\B j} }  &=& -2 \ g_{ij} 
                                                 \ \E_{\dot\A\dot\B} \ 
                                                 \bar Z 
\end{eqnarray}

All the other graded commutators vanish\footnote{The following
conventions concerning the internal $SU(2)_{R}$ symmetry
of $N=2$ supersymmetry have been used:
$Q_{\A}^{ \ i \ +} =  \bar Q_{\dot\A \ i}$,
$g^{12}= 1$,
$g^{ij} = -  g^{ji} = -g_{ij}$,
$g^{ij} g_{jk} = \delta^{i}_{ \ k}$}.                                       
Performing the change of basis the supersymmetry algebra in the 
new basis 
$D_{A} = \{ \partial_{m}, D_{\A}^{i},\bar D_{\dot\A j}, \partial_{z}, 
            \partial{\bar z} \}$ is

\begin{eqnarray}
\label{n=2susy_algebra2}
 \acom{ D_{\A}^{ \ i} }{ \bar D_{\dot\A j} }
           &=& -2 i \ \delta^{i}_{ \ j} \ \Sm_{ \ \A \dot\A}^{m} \ \partial_{m} 
\nonumber\\
 \acom{ D_{\A}^{ \ i} }{ D_{\B}^{ \ j} }  
                            &=& -2i \ g^{ij} \ \E_{\A\B} \ \partial_{z} 
\nonumber\\
\acom{ \bar D_{\dot\A i} }{ \bar D_{\dot\B j} } 
      &=& -2i \ g_{ij}  \ \E_{\dot\A\dot\B} \ \partial_{\bar z}
\end{eqnarray}

Note that $\partial_{\bar z} = - \ \partial_{z}^{+}$ holds in this conventions.
Again all the other graded commutators vanish. This follows from
the $N=2$ torsion constraints:

\begin{eqnarray}
\label{n=2_torsion_constraints}
  T_{\A\dot\A \ j}^{ \ \ m \ i} =  2 \ \Sm_{ \ \ \A \dot\A}^{m}  \ 
                                    \delta^{i}_{ \ j}  
\hspace{1cm}
  T_{\A\B}^{ \ \  z  i j}       =  2 \ \E_{\A\B} \ \ g^{ij}
\hspace{1cm}
  T^{\dot\A\dot\B \bar z  jk}   = 2 \ \E^{\dot\A\dot\B} \ \ g^{jk}
\end{eqnarray}

Here the index $z$ is used as an internal index and all other torsions
vanish. The grading of the generators follows from their index-structure: 
$| P_{m} |  = | m | =  | Z | =  | z | = 0 $
and
$|Q_{\A}^{ \ i} | = | \A | =  | \bar Q_{\dot\A j}| = | \dot\A | =  1 $.
By the use of the $N=2$ superspace coordinates 
$z^{M} =  (x^{m}, \theta^{\A}_{ \ i},\bar\theta^{\dot\A j}, z, \bar z)$  
an explicit representation of the supersymmetry
generators can be given with  
$\partial_{\A}^{ \ i } = \frac{\partial}{\partial \T^{\A}_{ \ i}} $
and 
$\bar\partial_{\dot\A j } = \frac{\partial}{\partial \Tb^{\dot\A j}} $:

\begin{eqnarray}
\label{n=2susy_generators}
 Q_{\A}^{ \ i} &=&  \partial_{\A}^{ \ i } 
                 - \Sm_{ \ \ \A \dot\A}^{m} \ P_{m} \ \Tb^{\dot \A i}
                 +  \ g^{ij} \ \E_{\A\B} \ \T^{ \B }_{ \ j} \ Z 
\nonumber\\
\bar Q_{\dot\A j} &=& - \ \bar\partial_{\dot\A j } 
                 + \T^{\A}_{ \ j } \ \Sm_{ \ \ \A \dot\A}^{m} \ P_{m} 
                 +  \ g_{jk} \ \E_{\dot\A\dot\B} \ \Tb^{\dot\B k} \ \bar Z                                         
\end{eqnarray}

A change of basis leads to the following representation of the
spinorial derivatives in superspace

\begin{eqnarray}
\label{n=2susy_derivatives}
 D_{\A}^{ \ i} &=&  \partial_{\A}^{ \ i } 
                 + \ i \ \Sm_{ \ \ \A \dot\A}^{m} \ \partial_{m} \ \Tb^{\dot \A i}
                 - \ i \ g^{ij} \ \E_{\A\B} \ \T^{ \B }_{ \ j} \ \partial_{z} 
\nonumber\\
\bar D_{\dot\A j} &=& - \ \bar\partial_{\dot\A j } 
       - \ i \ \T^{\A}_{ \ j } \ \Sm_{ \ \ \A \dot\A}^{m} \ \partial_{m} 
       + \ i \ g_{jk} \ \E_{\dot\A\dot\B} \ \Tb^{\dot\B k} \ \partial_{\bar z}                                         
\end{eqnarray}

These spinorial derivatives commute with the standard supersymmetry
generators $Q_{\A}^{ \ i}$

\begin{eqnarray}
\label{n=2susy_algebra2_basis_constraint}
 \acom{ D_{\A}^{ \ i} }{ \bar Q_{\dot\B j} }       \ = \
 \acom{ D_{\A}^{ \ i} }{ Q_{\B}^{j} }              \ = \
\acom{ \bar D_{\dot\A i} }{ \bar Q_{\dot\B j} } \ = \
\acom{ \bar D_{\dot\A i} }{ Q_{\B}^{ \ j} }     \ = \    0.                                        
\end{eqnarray}

Now the concept of intergration in $N = 2$ superspace can be introduced.
With the definition

\begin{eqnarray}
 \int \ d\theta^{\A}_{ \ i}    &=& 0
 \hspace{2cm}
  \int \ d\theta^{\A}_{ \ i}  \ \theta_{\A}^{ \ i}  \ = \ 1                                        
\end{eqnarray}

one finds the well known result that integration and differentiation
give the same result for Grassmann variables:

\begin{eqnarray}
 \int \ d\theta^{\A}_{ \ i}    &=& \frac{1}{4} \ D^{\A}_{ \ i}                                    
\end{eqnarray}

Analogous one finds the following identities in superspace

\begin{eqnarray}
\label{grassmann_integration}
\int \ d\bar\theta_{\dot\A \ i}  &=& - \frac{1}{4} \ \bar D_{\dot\A \ i}
\nonumber\\
 \int \ d^{2}\theta_{i}^{ \ j}  &=& \frac{1}{12} \ D^{\A}_{ \ i} D_{\A}^{ \ j}
\nonumber\\
 \int \ d^{2}\bar\theta_{ \ i}^{j}  &=& \frac{1}{12} \ 
                                  \bar D_{\dot\A}^{ \ j} D^{\dot\A}_{ \ i}.                                               
\end{eqnarray}

To work in a basis independent way a vielbein 
$e_{M}^{ \ A}(z)$ with $ A \sim (a, \A i, \dot\A i, z,  \bar z)$  
and its inverse can be introduced

\begin{eqnarray}
\label{vielbein_property}
 e_{M}^{ \ A} (z) \  e_{A}^{ \ N} (z) = \delta_{M}^{ \ N} 
\hspace{1cm}
 e_{A}^{ \ M} (z) \  e_{M}^{ \ B} (z) = \delta_{A}^{ \ B}
\end{eqnarray}

Using $D_{A} =  e_{A}^{ \ M} \ \frac{\partial}{\partial z^{M}}$ the
exterior derivative is $d = e^{A} D_{A}$.
In particular there is also an explicit representation of the vielbein

\begin{eqnarray}
\label{vielbein_matrix}
e_{A}^{ \ \ M} &=&
\left (
   \begin{array}{ccccc}
\delta_{a}^{ \ m} &  0 &  0 & 0 &  0                    \\
 \ i \ \Sm_{ \ \ \A \dot\A}^{m}  \ \Tb^{\dot \A i} & 
\delta_{\A}^{ \ \mu} \delta_{j}^{ \ i} & 0 &
i\ \E_{\A\B} \ \T^{ \B }_{ \ j}  \ g^{ji}   &  0    \\
i \ g^{ij} \ \T^{\A}_{ \ j } \ \Sm_{ \ \ \A \dot\B}^{m} \E^{\dot\B\dot\A} &
0 & \delta^{\dot\A}_{ \ \dot\mu} \delta_{j}^{ \ i} &
0 &  i \ \Tb^{\dot\A i}  \\
0 &  0 &  0 & \delta_{z}^{ \ z} &  0 \\
0 &  0 &  0 & 0 &  \delta_{\bar z}^{ \ \bar z} \\
    \end{array}
\right ) 
\nonumber\\
\end{eqnarray}

Its inverse is given in this specific basis as

\begin{eqnarray}
\label{inverse_vielbein_matrix}
e_{M}^{ \ \ A} &=&
\left (
   \begin{array}{ccccc}
\delta_{m}^{ \ a} &  0 &  0 & 0 &  0                    \\
- i \ \Sm_{ \ \ \mu \dot\mu}^{a}  \ \Tb^{\dot\mu j} & 
\delta^{\A}_{ \ \mu} \delta^{j}_{ \ i} & 0 &
i\ \E_{\mu\nu} \ \ g^{jk} \T^{ \nu }_{ \ k}    &  0    \\
- i \ g^{jk} \ \T^{\nu}_{ \ k }\Sm_{ \ \ \nu \dot\nu}^{a}\E^{\dot\nu\dot\mu} &
0 & \delta_{\dot\A}^{ \ \dot\mu} \delta^{j}_{ \ i} &
0 &  -i \ \Tb^{\dot\mu j}  \\
0 &  0 &  0 & \delta_{z}^{ \ z} &  0 \\
0 &  0 &  0 & 0 &  \delta_{\bar z}^{ \ \bar z} \\
    \end{array}
\right ) 
\nonumber\\
\end{eqnarray}

Moreover, in this basis there is torsion, which is in general defined
as the exterior derivative of the vielbein
$ d e^{A} =  i \ T^{A} = \frac{1}{2} \ e^{C} \ e^{B} \ i \ T_{BC}^{ \ \ A}$.
In our specific basis this leads to

\begin{eqnarray}
\label{vielbein_property_torsion_2}
 d e^{a}  &=&  e^{\A}_{ \ i} \ e^{\dot\A j} \ i \ 
               T_{\A\dot\A \ j}^{ \ \ a \ i}
            =   e^{\A}_{ \ i} \ e^{\dot\A j} \ 
               2 \ i \ \Sm_{ \ \ \A \dot\A}^{a}  \ \delta^{i}_{ \ j}  
\nonumber\\
 d e^{z}  &=&  \frac{1}{2} \ e^{\A}_{ \ i} \  e^{\B}_{ \ j} \
               \ i \ T_{\A\B}^{ \ \  z  i j}
            =  e^{\A}_{ \ i} \  e^{\B}_{ \ j} \
               \ i \ \E_{\A\B} \ \ g^{ij}
\nonumber\\
 d e^{\bar z} &=&  \frac{1}{2} \ e_{\dot\A j } \ e_{\dot\B k }   \
               \ i \ T^{\dot\A\dot\B \bar z  jk}
            =  e_{\dot\A j } \ e_{\dot\B k }   \
               \ i \ \E^{\dot\A\dot\B} \ \ g^{jk}
\end{eqnarray}


Let us now discuss the hypermultiplet in the framework of this
$N = 2$ superspace. The $8_{B} + 8_{F}$ hypermultiplet  

\begin{eqnarray}
 \phi_{i} 
 &\sim& ( \ A_{i} \ | \ \chi_{\A} \ , \ \bar\psi_{\dot\A} \ || \ F_{i} \ )
\end{eqnarray}

contains two complex scalars, two Weyl fermions and two complex auxiliary
fields - even if the central charge is complex. It obeys the following 
constraint:

\begin{eqnarray}
\label{constraint}
 D_{\A}^{ \ (i} \phi^{j)}           &=&  0
\hspace{2cm}
 \bar D_{\dot\A}^{ \ (i} \phi^{j)} \ = \ 0
\end{eqnarray}

In addition the following reality constraint must hold to eliminate superfluous
component fields if the central charge is complex:

\begin{eqnarray}
\label{reality_constraint}
  ( \partial_{z} \ + \ \partial_{\bar z} ) \ \phi_{i} &=& 0
\hspace{2cm}
   ( \partial_{z} \ + \ \partial_{\bar z} ) \ \bar\phi^{i} \ = \ 0
\end{eqnarray}

This constraint is the {\em only} possibility to extend the
hypermultiplet of Fayet and Sohnius [\ref{Fayet},\ref{Sohnius}]
to the case of a complex central charge without introducing new
component fields and preserving supersymmetry.
Hence the phase of the central charge is irrelevant for the
hypermultiplet of Fayet and Sohnius. 
Eq. (\ref{reality_constraint}) becomes trivial
if the central charge is real. Moreover (\ref{reality_constraint}) 
yields the following power expansion of the hypermultiplet in terms of
the internal coordinates $z$ and $\bar z$

\begin{eqnarray}
\label{reality_evolution}
  \phi_{i} &=& \sum_{k=0}^{\infty} \ \alpha_{k,i} (x,\theta,\bar\theta) \
                (z - \bar z)^{k}. 
\end{eqnarray}

In the following we will determine all coefficients
$\alpha_{k,i} (x,\theta,\bar\theta)$ and we will show that only the
first coefficients need to be considered in a 
renormalizable low-energy effective theory in four dimensions.


Using (\ref{constraint}) and the algebra one can calculate the following
superfield identities:

\vspace{0,3cm}

{\em Superfield identities for $\phi_{i}$:}

\begin{eqnarray}
 D_{\A}^{ \ i} \ \phi_{j} &=& \frac{1}{2} \ \delta^{i}_{ \ j}
                              \ D_{\A}^{ \ k} \ \phi_{k}
\nonumber\\
 \bar D_{\dot\A \ i} \ \phi_{j} &=& - \frac{1}{2} \ g_{ij}
                              \ \bar D_{\dot\A k} \ g^{kl} \ \phi_{l}
\nonumber\\                   
  D_{\A}^{ \ i} \ D_{\B}^{ \ j} \ \phi_{j} &=& - \ 4 \ i \ \E_{\A\B}
                                             \ g^{ij} \ \partial_{z} \ \phi_{j}                             
\nonumber\\
 \bar D_{\dot\A i} \ \bar D_{\dot\B j} \ g^{jk} \phi_{k} &=&
       - \ 4 \ i \ \E_{\dot\A\dot\B} \ \partial_{\bar z} \ \phi_{i}                             
\nonumber\\ 
 D_{\A}^{ \ i} \ \bar D_{\dot\B j} \ g^{jk} \phi_{k} &=&
     - \ 4 \ i \ \Sm^{m}_{ \ \A\dot\B} \ \partial_{m} \ g^{ij} \ \phi_{j}                             
\nonumber\\   
\bar D_{\dot\A i} \ D_{\B}^{ \ j} \ \phi_{j} &=&
     - \ 4 \ i \ \Sm^{m}_{ \ \B\dot\A} \ \partial_{m} \ \phi_{i}                             
\nonumber\\  
\partial_{\bar z} \ D_{\A}^{ \ i} \ \phi_{i} &=&
     -  \ \Sm^{m}_{ \ \A\dot\A} \ \partial_{m} \ \E^{\dot\A\dot\B} \ 
     \bar D_{\dot\B i} \ g^{ij} \ \phi_{j}                              
\nonumber\\
\partial_{z} \ \bar D_{\dot\A i} \ g^{ij} \ \phi_{j} &=&
     -  \ \Sm^{m}_{ \ \A\dot\A} \ \partial_{m} \ \E^{\A\B} \ 
     D_{\B}^{ \ i} \ \phi_{i}                              
\nonumber\\                                                       
\end{eqnarray}

{\em Superfield identities for $\bar\phi^{i}$:}

\begin{eqnarray}
 \bar D_{\dot\A i} \ \bar\phi^{j} &=& \frac{1}{2} \ \delta^{i}_{ \ j}
                              \ \bar D_{\dot\A k} \ \bar\phi^{k}
\nonumber\\
 D_{\A}^{ \ i} \ \bar\phi^{j} &=& - \frac{1}{2} \ g^{ij}
                              \ D_{\A}^{ \ k} \ g_{kl} \ \bar\phi^{l}
\nonumber\\                   
  D_{\A}^{ \ i} \ D_{\B}^{ \ j} \ g_{jk} \ \bar\phi^{k} &=& 
  - \ 4 \ i \ \E_{\A\B} \ \partial_{z} \ \bar\phi^{i}                             
\nonumber\\
 \bar D_{\dot\A i} \ \bar D_{\dot\B j} \ \bar\phi^{j} &=&
   - \ 4 \ i \ \E_{\dot\A\dot\B} \ g_{ij} \ \partial_{\bar z} \ \bar\phi^{j}                             
\nonumber\\ 
 D_{\A}^{ \ i} \ \bar D_{\dot\B j} \  \bar\phi^{j} &=&
     - \ 4 \ i \ \Sm^{m}_{ \ \A\dot\B} \ \partial_{m} \ \bar\phi^{i}                             
\nonumber\\
 \bar D_{\dot\A i} \ D_{\B}^{ \ j} \ g_{jk} \bar\phi_{k} &=&
   - \ 4 \ i \ g_{ij} \ \Sm^{m}_{ \ \B\dot\A} \ \partial_{m} \ \bar\phi^{j}                             
\nonumber\\
 \partial_{z} \ \bar D_{\dot\A i} \ \bar\phi^{i} &=&
     -  \ \Sm^{m}_{ \ \A\dot\A} \ \partial_{m} \ \E^{\A\B} \ 
     D_{\B}^{ \ i} \ g_{ij} \ \bar\phi^{j}                              
\nonumber\\
\partial_{\bar z} \  D_{\A}^{ \ i} \ g_{ij} \ \bar\phi^{j} &=&
     -  \ \Sm^{m}_{ \ \A\dot\A} \ \partial_{m} \ \E^{\dot\A\dot\B} \ 
     \bar D_{\dot\B i} \ \bar\phi^{i}                              
\nonumber\\                                                       
\end{eqnarray}

These identities yield the important result

\begin{eqnarray}
\label{PRE_BPS}
\Box \ \phi_{i}     &=& \partial_{z} \  \partial_{\bar z} \ \phi_{i}
\hspace{2cm}
\Box \ \bar\phi^{i} \ = \ \partial_{z} \ \partial_{\bar z} \ \bar\phi^{i}.                                                     
\end{eqnarray}

Hence for any massive hypermultiplet the BPS bound (\ref{BPS_bound}) 
is saturated on-shell.
Moreover we find by the use of (\ref{PRE_BPS}) for the coefficients
$\alpha_{k,i}$ of the power expansion (\ref{reality_evolution}) the 
following recursion formula

\begin{eqnarray}
\label{recursion}
   k \ (k-1) \ \alpha_{k,i} \ + \ \Box \ \alpha_{k-2,i} &=& 0
\hspace{2cm} k \geq 2 
\end{eqnarray}
 
As a consequence of (\ref{recursion}) we can construct the complete
power expansion of the hypermultiplet in superspace from the
coefficients $\alpha_{0,i}$ and $\alpha_{1,i}$. To do so  we
define the independent four-dimensional hypermultiplet component fields:

\begin{eqnarray}
\label{comp_level_hypermultiplet}
 \phi_{i |} &=& A_{i} \ (x)
\nonumber\\
 D_{\A}^{ \ i}  \phi_{i |} &=&  2 \sqrt{2} \ \chi_{\A} \ (x)
\nonumber\\
 \bar D_{\dot\A i} \ g^{ij} \ \phi_{j |} &=& -2 \sqrt{2} \ 
       \bar\psi_{\dot\A} \ (x)
\nonumber\\
 \partial_{z} \ \phi_{ i |} &=& F_{i} \ (x)
\nonumber\\
 \partial_{\bar z} \ \phi_{ i |} &=& - \ F_{i} \ (x) 
\end{eqnarray}
\begin{eqnarray}
\label{comp_level_hypermultiplet_2}
 \bar\phi^{i}_{ \ |} &=& \bar A^{i} \ (x)
\nonumber\\
\bar D_{\dot\A i}  \bar\phi^{i}_{ \ |} &=&  2 \sqrt{2} \
                                           \bar\chi_{\dot\A} \ (x)
\nonumber\\
 D_{\A}^{ \ i} \ g_{ij} \ \bar\phi^{i}_{ \ |} &=& 2 \sqrt{2} \ \psi_{\A} \ (x)
\nonumber\\
 \partial_{\bar z} \  \bar\phi^{i}_{ \ |} &=& - \ \bar F^{i} \ (x)
\nonumber\\
 \partial_{z} \  \bar\phi^{i}_{ \ |} &=& \bar F^{i} \ (x). 
\end{eqnarray}

The component fields transform under supersymmetry transformations
generated by the operator
$\delta = \xi^{\A}_{ \ i} D_{\A}^{ \ i} +
          \bar\xi_{\dot\A i} \bar D^{\dot\A i}$ as follows:

\begin{eqnarray}
\delta \  A_{i} &=& \sqrt{2} \ \xi^{\A}_{ \ i} \ \chi_{\A}  \ + \
                    \sqrt{2} \ \bar\xi_{\dot\A i} \ \bar\psi^{\dot\A}
\nonumber\\
\delta \ \chi_{\A} &=& - \ \sqrt{2} \ i \ \xi^{\B}_{ \ i} \ g^{ij} \ \E_{\B\A}
                       \ F_{j}   \ - \
                       \sqrt{2} \ i \ \bar\xi_{\dot\B i} \ g^{ij} \
                        \E^{\dot\B\dot\A} \
                       \Sm_{ \ \A\dot\A}^{m} \ \partial_{m} \ A_{j}
\nonumber\\
\delta \ \bar\psi_{\dot\A} &=& 
     \sqrt{2} \ i \ \xi^{\B}_{ \ i} \ g^{ij} \ 
     \Sm^{m}_{ \ \B\dot\A} \ \partial_{m} \ A_{j}
     \ - \
     \sqrt{2} \ \bar\xi_{\dot\A i} \ g^{ij} \ F_{j}
\nonumber\\
\delta \ F_{i} &=& - \ \sqrt{2} \ \xi^{\B}_{ \ i} \ 
                   \Sm^{m}_{ \ \B\dot\A} \ \partial_{m} \ \bar\psi^{\dot\A}
                   \ + \
                   \sqrt{2} \ \bar\xi_{\dot\B i} \ 
                   \E^{\dot\B\dot\A} \
                   \Sm_{ \ \A\dot\A}^{m} \ \partial_{m} \ \chi^{\A}
\end{eqnarray}

The transformations of the component fields with respect to the
central charge are generated by the operator
$\hat\delta = \xi^{z} \partial_{z} + \bar\xi^{\bar z} \partial_{\bar z}$:

\begin{eqnarray}
\hat\delta \  A_{i} &=&  ( \xi^{z} \ - \ \bar\xi^{\bar z} ) \ F_{i}
\nonumber\\
\hat\delta \ \chi_{\A} &=& ( \bar\xi^{\bar z} \ - \ \xi^{z} ) \
                       \Sm^{m}_{ \ \A\dot\A} \ \partial_{m} \ \bar\psi^{\dot\A}
\nonumber\\
\hat\delta \ \bar\psi_{\dot\A} &=& ( \xi^{z} \ - \ \bar\xi^{\bar z} ) \
                       \Sm_{ \ \A\dot\A}^{m} \ \partial_{m} \ \chi^{\A}
\nonumber\\
\hat\delta \ F_{i} &=& ( \bar\xi^{\bar z} \ - \ \xi^{z} ) \ \Box \ A_{i}
\end{eqnarray}

Thus the central charge of the hypermultiplet vanishes 
for a massless free field theory if the equations of motion hold.
\\
From the definition of the independent component fields of the hypermultiplet
and the supersymmetry algebra we find
the two coefficients $\alpha_{0,i}$ and $\alpha_{1,i}$ 
of the power expansion (\ref{reality_evolution}):

\begin{eqnarray}
\alpha_{0,i} &=& A_{i} \ + \ \sqrt{2} \ \theta^{\A}_{ \ i} \ \chi_{\A}
                 \ - \ \sqrt{2} \ \bar\theta_{\dot\A i} \ \bar\psi^{\dot\A}
  \ + \ 3 \ i \ \theta^{\A}_{ \ j} \ \Sm^{m}_{ \ \A\dot\A} \ \partial_{m} \
  \bar\theta^{\dot\A j} \ A_{i} 
\nonumber\\ & &
  \ + \ \frac{3}{2} \ i \ \theta^{\A}_{ \ j} \ g^{jk} \ \E_{\A\B} \ 
  \theta^{\B}_{ \ k} \ F_{i} \ 
  \ + \ \frac{3}{2} \ i \ \bar\theta_{\dot\A j} \ g^{jk} \ \E^{\dot\A\dot\B} \ 
  \bar\theta_{\dot\B k} \ F_{i} \ + \ \cdots
\\
\nonumber\\
\alpha_{1,i} &=& F_{i} 
  \ - \ \sqrt{2} \ \theta^{\A}_{ \ i} \ \Sm^{m}_{ \ \A\dot\A} \ \partial_{m} \
  \bar\psi^{\dot\A}
  \ - \ \sqrt{2} \ \bar\theta_{\dot\A i} \ \E^{\dot\A\dot\B} \
  \Sm^{m}_{ \ \B\dot\B} \ \partial_{m} \ \chi^{\B}
\nonumber\\ & &
  \ + \ 3 \ i \ \theta^{\A}_{ \ j} \ \Sm^{m}_{ \ \A\dot\A} \ \partial_{m} \
  \bar\theta^{\dot\A j} \ F_{i}
  \ - \ \frac{3}{2} \ i \ \theta^{\A}_{ \ j} \ g^{jk} \ \E_{\A\B} \ 
  \theta^{\B}_{ \ k} \ \Box \ A_{i} \
\nonumber\\ & &
  \ + \ \frac{3}{2} \ i \ \bar\theta_{\dot\A j} \ g^{jk} \ \E^{\dot\A\dot\B} \ 
  \bar\theta_{\dot\B k} \ \Box \ A_{i} \ + \ \cdots. 
\end{eqnarray}

The dots stand for a finite series in Grassmann variables that is not 
needed in the following. This technique to determine the dependence 
on the central charge can be applied to any supersymmetric multiplet.
Note that this expansion contains derivatives of the component field
$F_{i}$. The dimension of the coefficients
$\alpha_{k,i}$ increase linear in $k$: dim $(\alpha_{k,i}) = k + 1$.
Thus it is not necessary to consider
the full hypermultiplet in a renormalizable low-energy effective theory.
By definition the low-energy effective theory is
of second order in derivatives in the bosonic fields and
of first order in the fermionic fields.
Hence, to find the free action of a massive hypermultiplet
in a low-energy effective theory in four dimensions it is enough
to consider the approximation

\begin{eqnarray}
\label{LE_expansion}
\phi_{i} &\approx& \alpha_{0,i} (x,\theta,\bar\theta) \ + \ 
                  \alpha_{1,i}  (x,\theta,\bar\theta) \ (z - \bar z)  \ + \ 
                  \frac{1}{2} \ \Box \ A_{i} (x) \ (z - \bar z)^{2}.
\end{eqnarray}

Using (\ref{LE_expansion}) the Lagrangian of the hypermultiplet
in $N = 2$ superspace is 

\begin{eqnarray}
\label{action}
\lagrange &=& - \ \frac{1}{2} \ \int d^{2}\theta_{i}^{ \ j} \ L_{j}^{ \ i}
              (\phi,\bar\phi)
              \ + \
\frac{1}{2} \ \int d^{2}\bar\theta_{ \ i}^{j} \  L_{j}^{ \ i}
              (\phi,\bar\phi)               
\end{eqnarray}

with the `kernel'

\begin{eqnarray}
 L_{j}^{ \ i} (\phi,\bar\phi) &=&
     \phi_{j} \ (m \ + \  i \ \partial_{\bar z}) \ \bar\phi^{i} \ + \
     \bar\phi^{i} \ (m \ + \  i \ \partial_{z}) \ \phi_{j}.
\end{eqnarray}

Projection to the lowest component $(\theta = \bar\theta = z =  \bar z =  0)$
(\ref{action}) yields up to total derivatives

\begin{eqnarray}
\label{action_2}
\lagrange &=& - \ \partial^{m} \ \bar A^{i} \ \partial_{m} \ A_{i} 
    \ - \
    i \ \chi^{\A} \ \Sm^{m}_{\ \A\dot\A} \ \partial_{m} \ \bar\chi^{\dot\A} 
    \ - \
    i \ \psi^{\A} \ \Sm^{m}_{\ \A\dot\A} \ \partial_{m} \ \bar\psi^{\dot\A}
    \ + \ 
    \bar F^{i} F_{i}
\nonumber\\ & &
    \ + \ m \ \psi^{\A} \  \chi_{\A} 
    \ + \ \ m \ \bar\psi_{\dot\A} \  \bar\chi^{\dot\A}
    \ - \ i \ m \ A_{i} \ \bar F^{i} \ + \ i \ m \ \bar A^{i} \ F_{i}.
\end{eqnarray}

Eliminating the auxiliary fields via their equations of motion
$\bar F^{i} = - i m \bar A^{i}$ and $F_{i} = i m A_{i}$
we find the Lagrangian of a massive hypermutliplet in a free field
theory. If the equation of motion hold we have for a massless 
hypermultiplet 

\begin{eqnarray}
  (\hat\delta \phi_{i})_{\mbox{on-shell}}&=& 0.
\end{eqnarray}

Thus, the power expansion (\ref{reality_evolution}) breaks down
and the expansion of a massless hypermultiplet
in $N=2$ superspace is simply

\begin{eqnarray}
\phi_{i \ \mbox{on-shell}} &=& A_{i} \ + \
 \sqrt{2} \ \theta^{\A}_{ \ i} \ \chi_{\A}
 \ - \ \sqrt{2} \ \bar\theta_{\dot\A i} \ \bar\psi^{\dot\A}
 \ + \ 3 \ i \ \theta^{\A}_{ \ j} \ \Sm^{m}_{ \ \A\dot\A} \ \partial_{m} \
 \bar\theta^{\dot\A j} \ A_{i}
\nonumber\\ 
\end{eqnarray}

To take only the lowest component of (\ref{action}) is only allowed
if the higher components of the Lagrangian are total derivatives, which
don't contribute to the action. This is equivalent to the fact that the
action is invariant under supersymmetry and central charge 
transformations\footnote{By the use of the algebra the action is invariant
under central charge transformations if it is invariant under supersymmetry
transformations. For convenience we consider both cases.}. To show this 
it is sufficient to study the following superspace expression:

\begin{eqnarray}
\label{dummy_action}
\tilde\lagrange  \ = \ \int d^{2}\theta_{i}^{ \ j} \
                    \phi_{j} \ \bar\phi^{i}
                    \ - \ \int d^{2}\bar\theta_{ \ i}^{j} \  
                    \phi_{j} \ \bar\phi^{i}
 &=&
- \ i \ \phi_{i} \ \partial_{\bar z} \ \bar\phi^{i} \ - \
 i \ \bar\phi^{i} \ \partial_{z} \ \bar\phi^{i} \ + \
\nonumber\\ & &
\frac{1}{8} \ D_{\A}^{ \ i} \ g_{ij} \ \bar\phi^{j} \ \E^{\A\B} \
D_{\B}^{ \ k} \ \phi_{k} \ + \
\nonumber\\ & &
\frac{1}{8} \ \bar D_{\dot\A \ i} \ g^{ij} \ \phi_{j} \ \E^{\dot\A\dot\B} \
\bar D_{\dot\B k} \ \bar\phi^{k} 
\nonumber\\
\end{eqnarray}

Using the superspace identities for the hypermutiplet it can be shown that
$\delta \ \tilde\lagrange = \partial_{m} \ K^{m}$ and
$\hat\delta \ \tilde\lagrange \ = \ \partial_{m} \ \hat K^{m}$
with

\begin{eqnarray}
K^{m} &=& \xi^{\A}_{ \ j} \ \Sm^{m}_{ \ \A \dot\A} \ \E^{\dot\A\dot\B} \
\frac{i}{2} \
\left \{
 \phi_{i} \ g^{ij} \ \bar D_{\dot\B k} \ \bar\phi^{k}  \ - \
 \bar\phi^{j} \ \bar D_{\dot\B \ k} \ g^{kl} \ \phi_{l} 
\right \} \ + \
\nonumber\\ & &
\bar\xi_{\dot\A i} \ \bar\Sm^{m \dot\A \B} \ \frac{i}{2} \
\left \{
 g^{ij} \ \phi_{j} \  D_{\B}^{ \ k} \ g_{kl} \ \bar\phi^{l} \ + \
 \bar\phi^{i} \ D_{\B}^{ \ k} \ \phi_{k}
\right \}
\nonumber\\
\hat K^{m} &=& ( \bar\xi^{\bar z} - \xi^{z} ) \ 
\ \frac{1}{8} \ \bar\Sm^{m \dot\B \B}
\left \{
  D_{\B}^{ \ i} \ \phi_{i} \ \bar D_{\dot\B j} \ \bar\phi^{j} \ - \
  \bar D_{\dot\B \ k} \ g^{kl} \ \phi_{l} \ 
  D_{\B}^{ \ i} \ g_{ij} \ \bar\phi^{j} 
\right \}
\end{eqnarray}

holds. Therefore the action

\begin{eqnarray}
S &=& \int d^{4}x \ \lagrange \ (z^{M})
\end{eqnarray}

does not depend on the superspace coordinates $z^{M}$
and no integration over the internal coordinate $z$ is needed. This is 
remarkable because (\ref{dummy_action}) and (\ref{action})
depends on all superspace coordinates and is no `highest component'.
The reason for this `phenomenon' is that for extended
supersymmetry it is possible to find invariants by integration
over subspaces of the full superspace provided the kernel
satisfies certain constraints [\ref{Cremmer},\ref{Townsend2}].
\\
Let us review this `phenomenon' here briefly: It is already well known from
$N = 1$ superspace that chiral kernels must be integrated over a 
chiral subspace to give the corresponding action. Furthermore 
in [\ref{Townsend2}] it has been
shown in the context of extended supersymmetry
that for kernels with no central charge a classification
for such superactions can be given. Moreover in [\ref{Cremmer}] it is shown
that there are kernels carrying central charge, which 
also only need an integration over a subspace of superspace to 
give a four dimensional action.
This is the situation considered here and we have discussed this 
`phenomenon' in the framework of a covariant $N=2$ superspace formulation. 
To be more concrete, if we take the central charge to be real and 
identify z with the coordinate of a fifth dimension, 
then (\ref{PRE_BPS}) becomes the equation
of motion of the hypermultiplet in five dimensions. Hence our 
Lagrangian (\ref{action}) depends on the coordinates of five dimensional
Minkowski space. The integration of this Lagrangian over four
dimensions gives an action, which is independent of the additional dimension
provided the five dimensional equation of motion of the 
hypermultiplet holds. Hence, if we
start with an on-shell theory in five dimensions
without central charge we can end up via dimensional reduction [\ref{Brink}]
with an off-shell theory in four dimensions
with central charge. This is the result of [\ref{Cremmer}].       
From the point of view of a higher dimensional theory
the generator of the complex central charge is related
to partial derivatives with respect to the internal coordinates.
For instance, one complex central charge corresponds to the momenta of the
two-dimensional internal space. Thus the constraint
(\ref{reality_constraint}), which eliminates the phase of the central
charge, means that the hypermultiplet depends only on one of the 
two internal coordinates. From the point of view of a four dimensional 
theory it is therefore enough to consider the Fayet-Sohnius 
hypermultiplet only with one central charge. To keep things more
general the starting point was, however, a four dimensional theory
with complex central charge. 
\\
\\

To conclude, the expansion of the hypermultiplet
in $N=2$ superspace has been given. In the whole discussion
the r\^ole of the central charge has been emphasized and
the connection to a low-energy effective theory has been discussed.
In the presence of a complex central charge a new constraint 
has been introduced to preserve supersymmetry and the field
content of the Fayet-Sohnius hypermultiplet. Then the superspace
action in the context of a free field theory has been given. 
It has been pointed out that the 
two-dimensional superspace measure projects the kernel
down to a very special subspace of superspace, which leaves the action
invariant under supersymmetry and central charge transformations, although
the Lagrangian depends on all superspace coordinates.
Therefore it is enough to consider the lowest component of the
Lagrangian, because the action only depends on this part.
\\
Finally the question arises for a matter coupled Yang-Mills theory.
The easiest way to derive such a model is to gauge the central charge
as in [\ref{Gaida}]. The expansion of the hypermultiplet
with respect to the internal coordinates $z$ and $\bar z$
gets additional contributions in this case but the general
construction remains the same. The symmetries of such an 
$N=2$ Super Yang-Mills model
are given by `special geometry' and `Hyperk\"ahler geometry' 
[\ref{Scalar_Manifolds}].

\vspace{0,3cm}


{\bf Acknowledgement:} I would like to thank
S. Ketov for helpful correspondence about the central charge and
D. L\"ust for reading the manuscript. Moreover I am happy to thank
C. Preitschopf and P. van Nieuwenhuizen for helpful discussions and critics.

%
%

\section*{References}
\begin{enumerate}
\item
\label{Montonen_Olive}
C. Montonen and D. Olive, Phys. Lett. {\bf 72B} (1977) 117.
{\em Magnetic monopoles as gauge particles ?}
\item
\label{Seiberg_Witten}
N. Seiberg and E. Witten, Nucl. Phys. {\bf B426} (1994) 19.
{\em Electric-magnetic duality, monopole condensation and confinement
in $N=2$ supersymmetric Yang-Mills theory}
\item
\label{Theisen_Klemm}
A. Klemm, W. Lerche, S. Theisen and S. Yankielowicz,
 Phys. Lett. {\bf 344B} (1995) 169.
{\em Simple singularities and $N=2$ supersymmetric Yang-Mills theory}
\item
\label{Louis}
B. de Wit, V. Kaplunovsky, J. Louis and D. L\"ust, 
Nucl. Phys. {\bf B451} (1995) 53.
{\em Perturbative couplings of vectormultiplets in $N=2$ heterotic string vacua}
\item
\label{Kachru_Vafa}
S. Kachru and C. Vafa,
 Nucl. Phys.  {\bf B450} (1995) 69.
{\em Exact results for $N=2$ compactifications of heterotic strings}
\item
\label{Luest}
G. Lopes Cardoso, D. L\"ust and T. Mohaupt,
Nucl. Phys.  {\bf B455} (1995) 131.
{\em Non-perturbative monodromies in $N=2$ heterotic string vacua}
\item
\label{Haag}
R. Haag, J.T. Lopusz\'anski and M. Sohnius, Nucl. Phys. {\bf B88} (1975) 257.
{\em All possible generators of supersymmetries of the S-matrix}
\item
\label{Fayet}
P. Fayet, Nucl. Phys. {\bf B113}  (1976) 135.
{\em Fermi-Bose Hypersymmetry}
\item
\label{Sohnius}
M. Sohnius, Nucl. Phys. {\bf B138} (1978) 109.
{\em Supersymmetry and central charges}
\\
M. Sohnius, Phys. Rep.  {\bf 128} (1985) 39.
{\em Introducing Supersymmetry}
\item
\label{Grimm_Wess_Sohnius}
R. Grimm, M. Sohnius and J. Wess, Nucl. Phys. {\bf B133} (1978) 275.
{\em Extended supersymmetry and gauge theories}
\item
\label{Gaida}
I. Gaida, Phys. Lett. {\bf 373B} (1996) 89.
{\em Extended supersymmetry with gauged central charge}
\item
\label{Witten_Olive}
D. Olive and E. Witten, Phys. Lett {\bf 78B} (1978) 97.
{\em Supersymmetry algebras that include topological charges}
\item
\label{Strominger}
A. Strominger, Nucl. Phys. {\bf B451} (1995) 96.
{\em Massless black holes and conifolds in string theory}
\item
\label{Kallosh}
R. Brooks, R. Kallosh and T. Ortin,
Phys. Rev. {\bf D52} (1995) 5797.
{\em Fermion zero modes and black-hole hypermultiplet with rigid 
supersymmetry}
\item
\label{Siegel}
N. Berkovits and W. Siegel, Nucl. Phys. {\bf B462} (1996) 213.
{\em Superspace effective actions for 4D compactifications of heterotic
and Type II superstring.}
\item
\label{harmonic_superspace}
A. Galperin, E. Ivanov, S. Kalitzin, V. Ogievetsky and E. Sokatchev,
Class. Quant. Grav. {\bf 1}  (1984) 469.
{\em Unconstrained $N=2$ matter, Yang-Mills and supergravity theories in 
harmonic superspace}
\item
\label{Townsend1}
P.S. Howe, K.S. Stelle and P.K. Townsend,
Nucl. Phys. {\bf B214} (1983) 519.
{\em The relaxed Hypermultiplet}
\item
\label{Lauwers}
B. de Wit, P.G. Lauwers and A. Van Proeyen,
Nucl. Phys. {\bf B255} (1985) 569.
{\em Lagrangians of $N=2$ Supergravity Matter Systems}
\item
\label{Wess_Bagger}
J. Bagger and J. Wess, Supersymmetry and Supergravity, Princeton
University Press.
\item
\label{Cremmer}
M. Sohnius, K.S. Stelle and P.C. West, Nucl. Phys. {\bf B173} (1980) 127.
{\em Dimensional reduction by legendr\'e transformation generates off-shell
 supersymmetric Yang-Mills theory}
\\
\\
E. Cremmer, S. Ferrara, K.S. Stelle and P.C. West,
Phys. Lett. {\bf B94} (1980) 349.
{\em Off-shell N = 8 supersymmetry with central charge}
\item
\label{Townsend2}
P.S. Howe, K.S. Stelle and P.K. Townsend,
Nucl. Phys. {\bf B191} (1981) 445.
{\em Superactions}
\item
\label{Brink}
L. Brink, J. Schwarz and J. Scherk, Nucl. Phys. {\bf B121} (1977) 77.
{\em Supersymmetric Yang-Mills theory}
\item
\label{Scalar_Manifolds}
B. de Wit, F. Vanderseypen and A. Van Proeyen,
Nucl. Phys. {\bf B400} (1993) 463.
{\em Symmetry structure of special geometries}
\\
\\
L. Andrianopoli, M. Bertolini, A. Ceresole, R.D. Auria,
S. Ferrara, P. Fr\'e and T. Magri, hep-th 9605032.
{\em $N=2$ Supergravity and $N=2$ Super Yang-Mills theory on general scalar
manifolds.}
\end{enumerate}

\end{document}